\PassOptionsToPackage{table,xcdraw}{xcolor}

\documentclass[sigconf]{acmart}

\copyrightyear{2024}
\acmYear{2024}
\setcopyright{rightsretained}
\acmConference[ICSE '24]{2024 IEEE/ACM 46th International Conference on Software Engineering}{April 14--20, 2024}{Lisbon, Portugal}
\acmBooktitle{2024 IEEE/ACM 46th International Conference on Software Engineering (ICSE '24), April 14--20, 2024, Lisbon, Portugal}\acmDOI{10.1145/3597503.3608138}
\acmISBN{979-8-4007-0217-4/24/04}

\usepackage{amsmath,amsfonts}
\allowdisplaybreaks

\usepackage{booktabs}
\usepackage{url}
\usepackage{graphicx}

\usepackage{paralist}
\usepackage{mdframed}
\usepackage{fancyvrb}

\usepackage[UKenglish]{babel}

\usepackage{framed}
\usepackage{graphicx}
\graphicspath{{./images/}}
\usepackage{placeins}
\usepackage{subcaption}
\usepackage[inline]{enumitem}
\usepackage{balance}
\usepackage{booktabs}  %
\usepackage{colortbl}

\usepackage[%
  group-minimum-digits=4,%
  list-final-separator={, and },%
  add-integer-zero=false,%
  free-standing-units,%
  binary-units,%
  detect-weight=true,%
  detect-inline-weight=math,%
]{siunitx}

\usepackage{microtype}

\usepackage{hyperref}

\usepackage[
    colorinlistoftodos,prependcaption,textsize=tiny
]{todonotes}

\usepackage{xspace}

\usepackage[capitalise]{cleveref}

\crefformat{footnote}{#2\footnotemark[#1]#3}

\usepackage{tabularx}

\usepackage{array}
\newcolumntype{L}[1]{>{\raggedright\let\newline\\\arraybackslash\hspace{0pt}}b{#1}}
\newcolumntype{C}[1]{>{\centering\let\newline\\\arraybackslash\hspace{0pt}}b{#1}}
\newcolumntype{R}[1]{>{\raggedleft\let\newline\\\arraybackslash\hspace{0pt}}b{#1}}

\newcommand{\multilinecellL}[1]{%
    \begin{tabular}{@{}l@{}}#1\end{tabular}
}
\newcommand{\multilinecellC}[1]{%
    \begin{tabular}{@{}c@{}}#1\end{tabular}
}

\newcommand{\summary}[2]{
    \vspace{1em}
    \noindent
    \colorbox{gray!20}{%
        \parbox{.97\linewidth}{%
                \textbf{Summary \textit{(#1)}}
            #2
        }%
    }%
}%

\usepackage{csquotes}
\usepackage{csvsimple}
\usepackage{listings}
\usepackage{multirow}
\usepackage[font=small]{caption}

\definecolor{codegreen}{rgb}{0,0.6,0}
\definecolor{codegray}{rgb}{0.5,0.5,0.5}
\definecolor{codepurple}{rgb}{0.58,0,0.82}
\definecolor{backcolour}{rgb}{0.95,0.95,0.92}

\lstdefinestyle{mystyle}{
  commentstyle=\color{codegreen},
  keywordstyle=\color{magenta},
  numberstyle=\tiny\color{codegray},
  stringstyle=\color{codepurple},
  basicstyle=\ttfamily\scriptsize,
  breakatwhitespace=false,
  breaklines=true,
  breakautoindent=true,
  captionpos=b,
  frame=single,
  keepspaces=true,
  numbers=left,
  numbersep=5pt,
  showspaces=false,
  showstringspaces=false,
  showtabs=false,
  tabsize=2,
  numberblanklines=true,
  showlines=true,
  xleftmargin=2em,
  framexleftmargin=2.2em
}

\let\origthelstnumber\thelstnumber
\makeatletter
\newcommand*\Suppressnumber{%
  \lst@AddToHook{OnNewLine}{%
    \let\thelstnumber\relax%
     \advance\c@lstnumber-\@ne\relax%
    }%
}

\newcommand*\Reactivatenumber[1]{%
  \setcounter{lstnumber}{\numexpr#1-1\relax}
  \lst@AddToHook{OnNewLine}{%
   \let\thelstnumber\origthelstnumber%
   \refstepcounter{lstnumber}
  }%
}

\makeatletter
\newcommand\notsotiny{\@setfontsize\notsotiny\@vipt\@viipt}
\makeatother

\newcommand{\linebreakand}{%
  \end{@IEEEauthorhalign}
  \hfill\mbox{}\par
  \mbox{}\hfill\begin{@IEEEauthorhalign}
}

\newcommand{\EvoSuite}{EvoSuite\xspace}
\newcommand{\EvoSuiteDefault}{EvoSuite\textsubscript{\textit{FS\-On}}\xspace}
\newcommand{\EvoSuiteFlaky}{EvoSuite\textsubscript{\textit{FS\-Off}}\xspace}
\newcommand{\NOD}{NOD\xspace}
\newcommand{\OD}{OD\xspace}

\begin{document}

\title{Do Automatic Test Generation Tools Generate Flaky Tests?}

\author{Martin Gruber}
\authornote{Both authors contributed equally to this research.}
\affiliation{%
  \institution{BMW Group, University of Passau}
  \city{Munich}
  \country{Germany}}
\email{martin.gr.gruber@bmw}

\author{Muhammad Firhard Roslan}
\authornotemark[1]
\affiliation{%
  \institution{University of Sheffield}
  \city{Sheffield}
  \country{United Kingdom}}
\email{mfroslan2@sheffield.ac.uk}

\author{Owain Parry}
\affiliation{%
  \institution{University of Sheffield}
  \city{Sheffield}
  \country{United Kingdom}}
\email{oparry1@sheffield.ac.uk}

\author{Fabian Scharnböck}
\affiliation{%
  \institution{University of Passau}
  \city{Passau}
  \country{Germany}}
\email{scharn05@ads.uni-passau.de}

\author{Phil McMinn}
\affiliation{%
  \institution{University of Sheffield}
  \city{Sheffield}
  \country{United Kingdom}}
\email{p.mcminn@sheffield.ac.uk}

\author{Gordon Fraser}
\affiliation{%
  \institution{University of Passau}
  \city{Passau}
  \country{Germany}}
\email{gordon.fraser@uni-passau.de}

\begin{abstract}

Non-deterministic test behavior, or flakiness, is common and dreaded among developers.
Researchers have studied the issue and proposed approaches to mitigate it.
However, the vast majority of previous work has only considered developer-written tests.
The prevalence and nature of flaky tests produced by test generation tools remain largely unknown.
We ask whether such tools also produce flaky tests and how these differ from developer-written ones.
Furthermore, we evaluate mechanisms that suppress flaky test generation.
We sample \num{6356} projects written in Java or Python.
For each project, we generate tests using \EvoSuite (Java) and Pynguin (Python), and execute each test \num{200} times, looking for inconsistent outcomes.
Our results show that flakiness is at least as common in generated tests as in developer-written tests.
Nevertheless, existing flakiness suppression mechanisms implemented in EvoSuite are effective in alleviating this issue (\SI{71.7}{\percent} fewer flaky tests).
Compared to developer-written flaky tests, the causes of generated flaky tests are distributed differently.
Their non-deterministic behavior is more frequently caused by randomness, rather than by networking and concurrency.
Using flakiness suppression, the remaining flaky tests differ significantly from any flakiness previously reported, where most are attributable to runtime optimizations and \EvoSuite-internal resource thresholds.
These insights, with the accompanying dataset, can help maintainers to improve test generation tools, give recommendations for developers using these tools, and serve as a foundation for future research in test flakiness or test generation.
\end{abstract}



\maketitle

\newcommand{\numPythonProjsGenSucc}{\num{4454}\xspace}
\newcommand{\numPythonTestsGenSucc}{\num{442338}\xspace}
\newcommand{\numJavaProjsGenSucc}{\num{1902}\xspace}
\newcommand{\numJavaTestsGenSucc}{\num{737498}\xspace}

\newcommand{\JavaEvosuiteDefaultTests}{\num{310193}\xspace}
\newcommand{\PythonDeveloperWrittenTests}{\num{303711}\xspace}
\newcommand{\JavaEvosuiteNoflakinesssuppressionTests}{\num{264000}\xspace}
\newcommand{\JavaDeveloperWrittenTests}{\num{163305}\xspace}
\newcommand{\PythonGeneratedTests}{\num{138627}\xspace}

\section{Introduction}%
\label{sec:introduction}

A {\it flaky test} is a test case that produces inconsistent results, meaning that the same test can pass or fail for no apparent reason, even when the system being tested has not changed~\cite{Parry2021}.
They are a major problem for software developers because they limit the efficiency of testing, complicate continuous integration, and reduce productivity~\cite{Lam2020a,Durieux2020,Micco2016}.
The negative effects of flaky tests are ubiquitous, experienced by large companies such as Google, Microsoft, and Facebook, as well as the developers of smaller open-source projects~\cite{Lam2019a,Machalica2019,Memon2017,Durieux2020}.
Indeed, recent surveys found that a majority of developers observe flaky tests on at least a monthly basis%
~\cite{Parry2022,gruber2022survey}.
As well as being a burden on developers, flaky tests are also a persistent problem in research, limiting the deployment of several state-of-the-art techniques for test selection and prioritization~\cite{Peng2020,Machalica2019,Yu2019}.

Increasing research interest in the area of flaky tests has produced a range of empirical studies regarding the causes, origins, and impacts of developer-written flaky tests~\cite{Luo2014,Vahabzadeh2015,Eck2019,Lam2020d}.
However, far less attention has been paid to flaky tests produced by automatic test generation tools~\cite{Shamshiri2015,Paydar2019}.
This research gap is problematic for several reasons.
Firstly, there is minimal guidance for developers regarding the sorts of flaky tests they might expect to receive from test generation tools and more crucially how to avoid them.
This threatens to detract from the positive benefits of such tools on the software development lifecycle as previously established~\cite{shamshiri2018automatically}.
Similarly, the maintainers of test generation tools have only limited information on the prevalence of automatically generated flaky tests and what causes them.
These insights are crucial for maintainers to prevent their tools from producing flaky tests.
Furthermore, researchers in the field of flaky tests would benefit from an analysis of how the root causes of developer-written flaky tests compare to those that are automatically generated.
Such an investigation would inform researchers on whether generated flaky tests are representative of their developer-written counterparts.
This would be useful for augmenting existing datasets of developer-written flaky tests with generated flaky tests~\cite{InternationalDatasetofFlakyTests}.

In this study, we used the popular search-based test generation tools EvoSuite~\cite{Fraser2011} and Pynguin~\cite{Lukasczyk2020} to generate test
suites for \num{1902} Java projects and \num{4454} Python projects respectively.
We repeatedly executed both the developer-written and automatically generated test suites of all the projects, consisting of nearly a million individual test cases, \num{200} times each to detect flaky tests.
We compared the prevalence of flakiness between both types of test suites and went one step further by comparing root causes, following our manual analysis on a random sample of \num{481} non-order-dependent flaky tests.
Furthermore, we performed the first scientific evaluation on the effectiveness of EvoSuite's built-in flaky test suppression feature.
Chiefly among our findings, we found that flaky tests are at least as common in generated tests as they are in developer-written tests, that EvoSuite's flaky test suppression feature can reduce the number of generated flaky tests by \SI{71.7}{\percent}, and that the distribution of the root causes of generated flaky tests differs to that of developer-written tests.

The main contributions of this study are as follows:

\noindent {\bf Contribution 1: Empirical study}: Our empirical study involving \num{6356} open-source projects is the largest among all previous studies on both flaky tests and search-based test generation.
Our study is also the first to analyze the root causes of automatically generated flaky tests.
See \cref{sec:Methodology} for more information.

\noindent {\bf Contribution 2: Recommendations}: The results of our study have important implications for software developers, maintainers of test generation tools, and researchers in the area of flaky tests.
From these, we are able to offer insights and recommendations that are actionable by these stakeholders.
See Sections \ref{sec:results} and \ref{sec:discussion} for more information.

\noindent {\bf Contribution 3: Dataset}: The dataset we collected for this study is the first publicly available dataset of flaky tests that contains automatically generated flaky tests and features a large manually annotated sample.
See our replication package for more information~\cite{GeneratedFlakinessDataset}.

\section{Background}%
\label{sec:background}

\subsection{Flaky Tests}
\label{sec:flaky_test_background}

Luo et al.~\cite{Luo2014} performed one of the earliest empirical studies of test flakiness.
They categorized the cause of the flaky tests repaired by developers in 201 commits across 51 open-source projects using the following ten categories:

\noindent {\bf 1. Async. Wait.} Test makes an asynchronous call but does not properly wait for it to finish, leading to intermittent failures.

\noindent {\bf 2. Concurrency.} Test spawns multiple threads that behave in an unsafe or unanticipated manner, such as a race condition.

\noindent {\bf 3. Floating Point.} Test uses floating points and is flaky due to unexpected results such as non-associative addition.

\noindent {\bf 4. Input/Output (I/O).} Test uses the filesystem and is flaky due to intermittent issues such as storage space limitations.

\noindent {\bf 5. Network.} Test depends on the availability of a network and is flaky when the network is unavailable or busy.

\noindent {\bf 6. Order Dependency.} Test depends on a shared value or resource that is modified by other test cases as a side effect.

\noindent {\bf 7. Randomness.} Test involves random number generators and is flaky due to not setting seeds, for example.

\noindent {\bf 8. Resource Leak.} Test does not release acquired resources (e.g. database connection) inducing flaky failures for itself or for other tests that require the same resources.

\noindent {\bf 9. Time.} Test relies on measurements of date and/or time.
Flakiness is caused by, for instance, discrepancies in precision and representation of time across libraries and platforms.

\noindent {\bf 10. Unordered Collection.} Test assumes a deterministic iteration order for an unordered collection-type object, such as a {\tt set}, leading to intermittent failures.

Eck et al.~\cite{Eck2019} asked Mozilla developers to categorize the causes of 200 flaky tests they had previously repaired.
The developers used the categories introduced by Luo et al. but with the option to create new categories if needed.
Following this, Eck et al. identified the following additional four categories:

\noindent {\bf 11. Too Restrictive Range.} Test includes a range-based assertion that excludes a portion of the valid output values.

\noindent {\bf 12. Test Case Timeout.} Test intermittently exceeds a pre-defined upper limit on its execution time.

\noindent {\bf 13. Platform Dependency.} Test outcome varies across the project's target platforms.

\noindent {\bf 14. Test Suite Timeout.} Test intermittently exceeds a pre-defined upper limit on the execution time of the test suite.

While studying flakiness in Python tests, Gruber et al.~\cite{Gruber2021} identified another root cause:

\noindent {\bf 15. Infrastructure.} Test fails intermittently due to issues outside the project code, but inside the execution environment (the container or the local host), for example, permission errors or lack of disk space.

\subsection{Automatic Test Generation}
\label{sec:automatic_test_generation_background}

{\bf EvoSuite} is an automatic test generation tool for Java projects~\cite{Fraser2011,Fraser2013}.
It uses a search-based approach to generate test suites that cover as much of the code under test as possible.
The tool is based on evolutionary algorithms, which means that it uses principles from biological evolution to search for test cases~\cite{McMinn2004}.
The search is guided by a fitness function that can be configured to optimize test suite generation for high line, branch, or mutation coverage.
The tool applies techniques from mutation testing to minimize the number of assertions in the generated tests~\cite{Yue2011}.
This ensures that they are more easily understandable to developers and not overly brittle.
A large-scale empirical study demonstrated that EvoSuite can achieve an average of 71\% branch coverage per class~\cite{Fraser2014}.
EvoSuite also provides ways to suppress ``unstable'' (flaky) tests, addressing this issue both during the evolutionary search process and after its completion.
EvoSuite detects unstable tests by controlling the environmental dependencies using bytecode instrumentation, resetting the state of static variables before executing each test, using mocks to replace non-deterministic calls, and compiling and executing the generated tests, removing failing tests~\cite{Arcuri2014}.

{\bf Pynguin} is an automatic test generation tool for Python projects~\cite{Lukasczyk2020,lukasczyk2022pynguin,lukasczyk2023empirical}.
Target languages aside, Pynguin and EvoSuite share several similarities.
Both tools use a search-based approach to generate test cases and both apply mutation testing to generate assertions.
Pynguin faces the additional challenge that Python is a dynamically typed language, meaning that generating test inputs of the appropriate type is much harder.
Therefore, Pynguin relies on existing type hints in function and class definitions in the code under test.
From these, Pynguin can apply type inference to attempt to determine the types of variables without hints.
A previous empirical evaluation of Pynguin found that it was able to achieve a mean branch coverage of \SI{71.6}{\percent} on 163 Python modules from 20 open-source projects~\cite{lukasczyk2023empirical}.

\section{Methodology}%
\label{sec:Methodology}

With our study, we aim to answer the following research questions:

\begin{description}
    \item[RQ1 (Prevalence):] How prevalent is flakiness in tests that were generated without flakiness suppression mechanisms?
    \item[RQ2 (Flakiness Suppression):] How many flaky tests can EvoSuite's flakiness suppression mechanism prevent?
    \item[RQ3 (Root Causes):] How do the root causes of generated flaky tests differ from those of developer-written tests?
\end{description}

\begin{figure*}
    \centering
    \includegraphics[width=\linewidth]{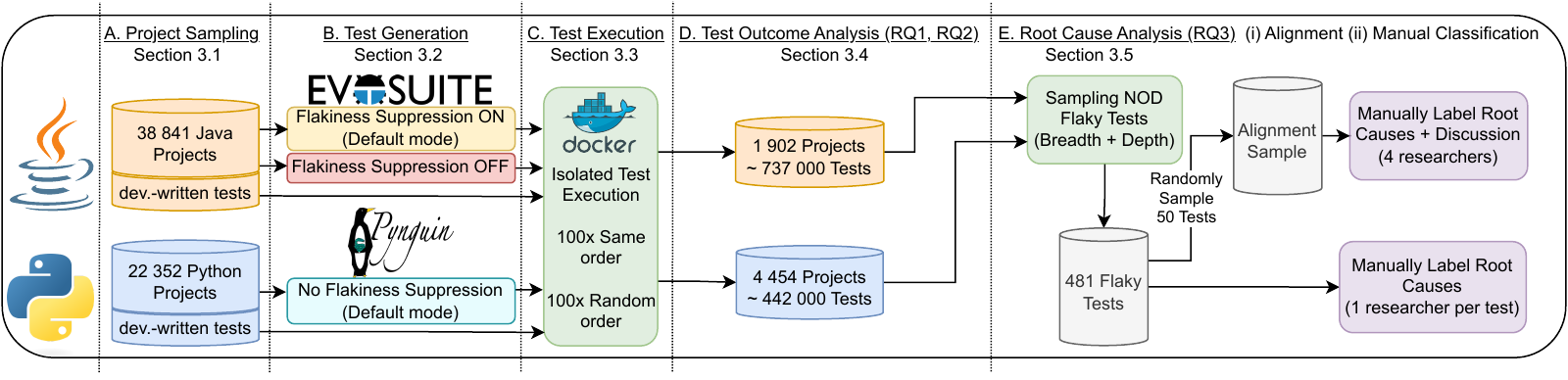}
    \caption{Study setup}%
    \label{fig:study_setup}
\end{figure*}

\cref{fig:study_setup} depicts an overview of our study setup.

\subsection{Project Sampling}%
\label{sec:project_sampling}

To collect subjects for our empirical study, we randomly sampled open-source projects written in Java and Python.
These are two of the most popular programming languages, which have also been the main targets for both test flakiness and test generation research.

\subsubsection{Java}%
\label{sec:java_project_sampling}

To collect Java projects, we used the index of the Maven Central Repository~\cite{MavenCentralRepository}, one of the official software repositories for Java.
The index is updated weekly for newly added projects or patches.
It consists of roughly \num{520000} unique packages (as of 2022--10--26). %
We iterated over the entire index and each project's Project Object Model (POM) to fetch the URL to the project repository.
We limited our project sampling to only include projects whose source code is available on GitHub and which use Maven as a build tool.
Since the project's POM in the Maven Central Repository does not include details of the build automation tool it is using---which means that the index includes projects built from Gradle or Ant---we crawled through the GitHub URLs to filter for projects that include a \texttt{pom.xml} file in the root of their repository.
This is to confirm that the project is using Maven as its build automation tool.
In total, we found \num{38841} Maven projects that include a link to the project repository on GitHub.
While some projects may lack developer-written tests, we did not exclude them during this crawling process (but we did so later, during the test outcome analysis).
We decided against sampling projects from existing flakiness databases, such as IDoFT~\cite{InternationalDatasetofFlakyTests}, because we did not want to limit our study to projects that already contain developer-written flaky tests.

\subsubsection{Python}%
\label{sec:python_project_sampling}

To collect Python projects, we used the dataset from Gruber et al.~\cite{Gruber2021}, who studied flakiness in Python.
It consists of \num{22352} projects that were randomly sampled from the Python Package Index (PyPI)~\cite{pypi}, the official third-party software repository of Python.
Each project contains at least one test that could be executed using pytest~\cite{pytest} and its source code is available on GitHub.
Unlike IDoFT~\cite{InternationalDatasetofFlakyTests}, these projects do not all contain flaky tests:
The original study found \num{7571} flaky tests among \num{1006} projects.

\subsection{Test Generation}%
\label{sec:test_generation}

To generate tests for the sampled projects, we use state-of-the-art test generation tools for the respective language.
For Java, we use \EvoSuite~\cite{Fraser2011}, an automated test generation tool that utilizes meta-heuristic techniques to generate JUnit test suites.
For Python, we use Pynguin~\cite{Lukasczyk2020, lukasczyk2022pynguin, lukasczyk2023empirical}, a test generation tool that produces unit tests for Python programs.
The tests are generated from scratch, not relying on any existing developer-written tests as input~\cite{Rojas2016}.

\subsubsection{Java}%
\label{sec:java_test_generation}
We use the latest release of \EvoSuite at the time (v1.2.0) to generate tests for each Java project.
Since \EvoSuite applies multiple techniques to avoid creating flaky tests, and we want to measure the impact of these flakiness suppression mechanisms, we apply \EvoSuite under two configurations:
\textit{with} and \textit{without} Flakiness Suppression, which we refer to as \EvoSuiteDefault and \EvoSuiteFlaky.
The flakiness suppression parameters are turned on by default, which means that for \EvoSuiteDefault, we do not change any of the \EvoSuite parameters.
To generate tests without the flakiness suppression mechanisms (\EvoSuiteFlaky), we update several parameters of \EvoSuite that we extracted from previous studies~\cite{Fraser2013,Arcuri2014,Fraser2018}, and which we confirmed via in-depth discussions with one of the \EvoSuite maintainers on our team.
\cref{tab:evosuiteflaky-parameters} shows the parameters that we changed to deactivate EvoSuite's flakiness suppression.
These parameters consist of actions carried out during the test generation process to mitigate non-determinism.
They address factors related to managing environmental dependencies, establishing a virtual file system, mocking non-deterministic output such as Random \cite{randomJavaClass} and Calendar \cite{calendarJavaClass} classes, and resetting the state of static and final fields to avoid creating dependencies on other tests.
For each of the Java projects, both \EvoSuiteDefault and \EvoSuiteFlaky generate tests for every testable class in the system under test (SUT).
A class is considered testable if it has at least one public method.
EvoSuite aims to generate a test suite that covers all public methods.
We set the search budget for generating tests to two minutes per class for both \EvoSuiteDefault and \EvoSuiteFlaky, which is what previous tool competitions have used~\cite{Vogl2021,Schweikl2022}.

\begin{table}
    \centering
    \caption{\raggedright Updated parameters to deactivate flakiness suppression mechanisms (\EvoSuiteFlaky)}
    \label{tab:evosuiteflaky-parameters}
    \footnotesize
    \rowcolors{2}{gray!10}{white}
    \begin{tabularx}{\linewidth}{|l|X|p{0.48cm}|}
        \hline
        \textbf{Parameter}        & \textbf{Description}                                         & \textbf{New Value} \\ \hline
        Test Scaffolding          & Generate separate scaffolding file to execute the tests      & false          \\
        No Runtime Dependency     & Avoid runtime dependencies in JUnit test                     & true           \\
        JUnit Check               & Compile and run the resulting JUnit test suite               & false          \\
        Sandbox                   & \raggedright Executing tests in an independent testing environment        & false          \\
        Virtual FS                & Using virtual file system for all File I/O operations        & false          \\
        Virtual Net               & \raggedright Using virtual network for all network communications         & false          \\
        Replace Calls             & Replacing non-deterministic calls                            & false          \\
        Replace System In         & Replacing the InputStream mechanism with a mock              & false          \\
        Replace GUI               & Replacing the GUI calls with a mock                          & false          \\
        Reset Static Fields       & Call static constructors only after static field was changed & false          \\
        Reset Static Field Gets   & Call static constructors only after static field was read    & false          \\
        Reset Static Final Fields & Remove the static modified in target fields                  & false          \\ \hline
    \end{tabularx}
\end{table}

\subsubsection{Python}%
\label{sec:python_test_generation}

To generate tests for the Python projects, we use version 0.27.0 of Pynguin, the latest release at the time (2022--11-14).
Pynguin operates on module-level, and we apply it to each module contained in each of our sample projects.
Like Lukasczyk et al.~\cite{lukasczyk2023empirical}, we use a maximum search budget of ten minutes.
Unlike EvoSuite, Pynguin does not offer optional parameters for flakiness suppression.
Instead, it applies a re-execute-once strategy to \enquote*{filter out trivially flaky assertions, e.g., strings that include memory locations}~\cite{pynguinTrivialFlakyFilter}:
After generating a test that contains passing assertions, the test is executed again.
Any assertion that does not hold in this execution is excluded, as it was made on an apparently flaky value.
This behavior is inspired by EvoSuite's \texttt{JUnit Check}, however, since it is not optional and Pynguin has no further flakiness-relevant parameter, we execute Pynguin using only one configuration: the default settings.

Both \EvoSuite and Pynguin generate tests non-deterministically, meaning that they generate different test suites every time the tool is used.
Most of the previous studies investigating automatic test generation generated more than one test suite per project to take the random nature of the evolutionary search~\cite{Shamshiri2015,Lukasczyk2020} into account.
Unlike these, we generate only one test suite per class/module, since our study does not draw any conclusions by comparing individual projects or components.
Instead, we accommodate for the randomness in the test generation by sampling a large corpus of projects, which also contributes to the generalizability of our findings.

\subsection{Test Execution}%
\label{sec:test_execution}

To detect flaky tests, we execute all generated tests and all developer-written tests \num{100} times in the same order and \num{100} times in random orders.
This procedure follows other studies~\cite{Gruber2021,Lam2019} and allows us to distinguish order-dependent (OD) from non-order-dependent (NOD) flaky tests.
The test executions are conducted either directly through or inspired by FlaPy~\cite{gruber2023flapy}, a tool that allows re\-searchers to mine flaky tests from a given set of projects by repeatedly executing their test suites.
FlaPy ensures a fresh and isolated environment for each test execution and handles dependency installation.
Furthermore, it splits the runs into iterations, where each iteration is executed in a separate Docker container, which helps to avoid timeouts and detect environment issues, such as infrastructure flakiness~\cite{Gruber2021}.
In our case, we split the \num{200} runs into at least five iterations per project and test type.
Generated and developer-written tests are not executed together, but in separate iterations to avoid side-effects.

To install third-party dependencies of the project under evaluation, we use language-specific pipelines:
For Java this can easily be accomplished by using \texttt{`mvn dependency:copy-dependencies'} to make a copy of all the dependencies from the repository on our local machine.
We then update the environment variables to include all the dependencies that the project needs when executing the tests.
For Python this process is more complicated since the general landscape of build systems is more heterogeneous.
As we cannot rely on a standardized solution, we use FlaPy's built-in dependency installation heuristic which searches for \texttt{requirements.txt} (or similarly named) files and runs them against \texttt{pip}.
To execute the projects' test suites we use the JUnit Runner~\cite{junit4} for Java and pytest~\cite{pytest} for Python.
When conducting test executions in random orders, we shuffle the tests on class-level, which randomly sorts first the classes and then the tests within each class.
For Python, this can easily be accomplished using pytest's random-order plugin~\cite{PytestRandomOrderPlugin}.
For Java, we had to create a custom test runner since Maven's Surefire plugin~\cite{mavenSurefire} currently does not support this form of shuffling.

\subsection{Test Outcome Analysis}%
\label{sec:test_outcome_analysis_rq1_rq3}

\begin{table}
    \centering
    \caption{Size of our dataset}
    \label{tab:dataset_size}
\resizebox{\linewidth}{!}{
    \begin{tabular}{llcccc}
        \toprule
        &  & \# Projects & \multicolumn{3}{c}{\underline{\smash{\# Tests (\textit{FS} = Flakiness Suppression)}}} \\
        Language & \multilinecellL{Test Gen.\\Framework} & & \multilinecellC{Developer\\-written} & \multilinecellC{Generated\\Without \textit{FS}} & \multilinecellC{Generated\\With \textit{FS}}
            \\
        \midrule
        Java   & EvoSuite & \numJavaProjsGenSucc   & \JavaDeveloperWrittenTests   & \JavaEvosuiteNoflakinesssuppressionTests & \JavaEvosuiteDefaultTests \\
        Python & Pynguin & \numPythonProjsGenSucc & \PythonDeveloperWrittenTests & \PythonGeneratedTests                    &                           \\
        \bottomrule
    \end{tabular}
}
\end{table}

\begin{figure}[!tbp]
    \begin{subfigure}[]{0.23\textwidth}
        \includegraphics[width=\linewidth]{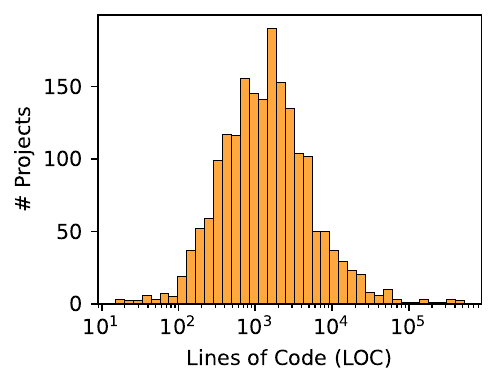}
        \caption{lines of code (Java)}
        \label{fig:java_loc_histogram}
    \end{subfigure}
    \hfill
    \begin{subfigure}[]{0.23\textwidth}
        \includegraphics[width=\linewidth]{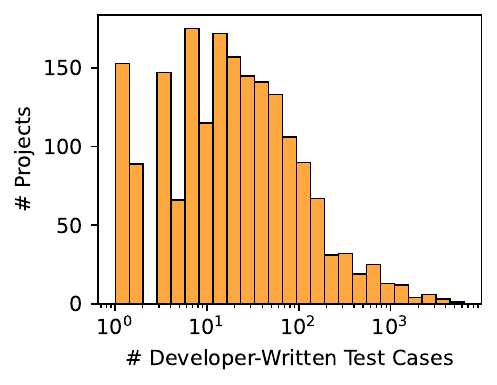}
        \caption{dev.-written Java tests / proj.}%
        \label{fig:java_number_of_test_cases}
    \end{subfigure}

    \vskip\baselineskip

    \begin{subfigure}[]{0.23\textwidth}
        \includegraphics[width=\linewidth]{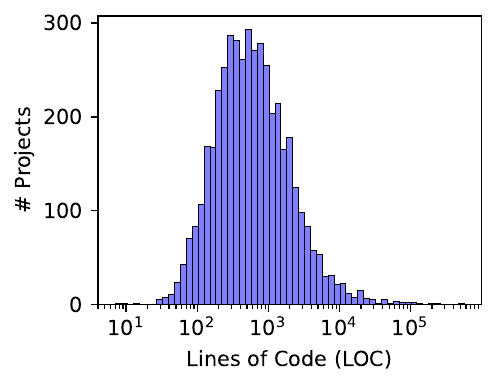}
        \caption{lines of code (Python)}
        \label{fig:python_loc_histogram}
    \end{subfigure}
    \hfill
    \begin{subfigure}[]{0.23\textwidth}
        \includegraphics[width=\linewidth]{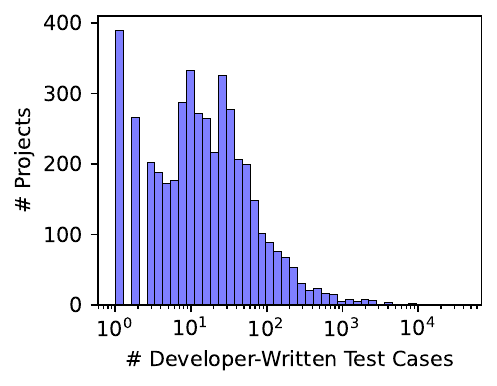}
        \caption{dev.-written Python tests / proj.}%
        \label{fig:python_number_of_test_cases}
    \end{subfigure}
    \caption{Dataset statistics}
    \label{fig:dataset_statistics}
\end{figure}

\cref{tab:dataset_size} depicts the number of projects for which we were able to successfully execute the developer-written tests, and successfully generate tests using EvoSuite or Pynguin.
We consider the execution of the developer-written test suite to be successful if at least one test case was executed without producing an \textit{Error} or \textit{Skip} outcome.
We consider the test generation to be successful if at least one executable test was generated.
Since we use the same generic setup processes for all projects of the same language, we were not able to successfully execute the developer-written test suite and the test generation for each sampled project.
Reasons for erroring test executions or test generation include project-specific requirements that go beyond standard third-party dependencies, such as setting global variables or installing system software.

To assert if we still derived a sufficiently large and diverse set of projects, we inspect two quantitative metrics:
the lines of source code (SLOC) of the projects measured via CLOC~\cite{adanial_cloc}, and
the number of developer-written tests they possess (\cref{fig:dataset_statistics}).
To assess if we applied the test generation tools properly, we look at the coverage that the generated tests achieved and compare it with the coverage reported by previous studies applying these tools.

\subsubsection{Java}%
\label{sec:java_test_outcome_analysis}
On average (mean) the Java projects possess \num{4948} lines of source code (median \num{1395}).
Only \SI{1.5}{\percent} of all projects contain less than \num{100} SLOC, whereas the largest project (\texttt{cosmos\--sdk-java}\footnote{\url{https://github.com/cloverzrg/cosmos-sdk-java}}) has more than \num{500000} lines of Java code.
The total number of SLOC is around 9.4 million for all projects combined.
\cref{fig:java_loc_histogram} shows the histogram of the SLOC distribution for the Java projects.
\cref{fig:java_number_of_test_cases} shows the number of developer-written test cases per Java project.
Multiple parametrizations of the same test case are treated as separate test cases, following a previous study~\cite{Gruber2021}.
In total, the projects possess \num{163305} developer-written tests and each project contains between \num{1} and \num{6315} test cases.
The median number of tests per project is \num{18}, and the mean is \num{85.9}.
As these figures about SLOC and test cases show, we have indeed derived a large and diverse sample of Java projects.
Both \EvoSuiteDefault and \EvoSuiteFlaky generated test suites with high branch- (over \SI{81}{\percent} mean) and line- (over \SI{84}{\percent} mean) coverage (\cref{fig:all_branchCoverage}).
This is similar to the reported code coverage of previous studies on \EvoSuite~\cite{Fraser2014,Fraser2015}.

\subsubsection{Python}%
\label{sec:Python_test_outcome_analysis}

\begin{figure}
    \includegraphics[width=\linewidth]{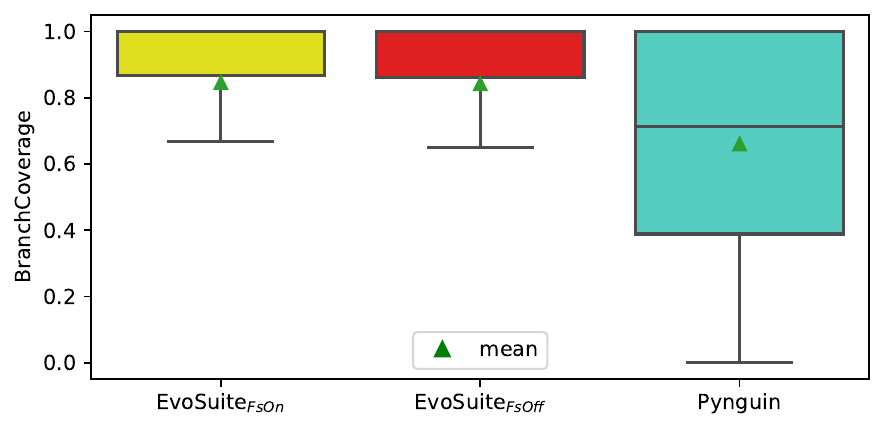}
    \caption{Branch coverage of generated tests}
    \label{fig:all_branchCoverage}
\end{figure}

\cref{fig:python_loc_histogram} depicts the size of the \numPythonProjsGenSucc Python projects in terms of SLOC.
The average (mean) project has \num{1755} SLOC (median \num{549.5}).
The largest project (\texttt{kuber}\footnote{\url{https://github.com/sernst/kuber}}) has more than \num{500000} lines of source code and only \SI{5.2}{\percent} of projects contain less than \num{100} SLOC.
Combined, the projects feature \num{7.8} million lines of Python code.
\cref{fig:python_number_of_test_cases} shows the size of the Python projects in terms of the number of developer-written tests they contain.
Like for the Java projects, the mean number of test cases per project (\num{68.2}) is substantially greater than the median (\num{14}), which is caused by a small number of very large projects.
In total, the Python projects contain \num{303711} developer-written test cases.
After inspecting the projects both in terms of number of SLOC and their number of test cases, we find no obvious bias towards overly small or large projects and conclude that we have derived a large and diverse sample of Python projects.
The Python tests generated by Pynguin yielded a mean branch coverage of \SI{66.0}{\percent} (\cref{fig:all_branchCoverage}) with a standard deviation of \SI{31.7}{}.
This performance is very similar to the one reached by Lukasczyk et al.~\cite{lukasczyk2023empirical}, the creators of Pynguin, who achieved a mean branch coverage of \SI{71.6}{\percent} with a standard deviation of \SI{30.5}{}.

\subsubsection{RQ1 (Prevalence)}%
\label{sec:rq1_prevalence}
To study the prevalence of flakiness in generated tests, we compare the number of flaky tests that were created by the test generation tools---without using flakiness suppression---to the number of flaky tests found in the developer-written tests of the respective language.
We regard a test as flaky if it yielded at least one passed and one failed or errored outcome~\cite{Gruber2021,dietrich2022flaky}.
Tests that switch between failing and erroring verdicts are therefore not considered as flaky, since both lead to a build failure and the test therefore does not contribute to the typical developer experience caused by flakiness (sporadically failing builds).
Furthermore, we look at the ratio between order-dependent flaky tests (which only show flaky behavior when run in random orders) and non-order-dependent flaky tests (which also show flaky behavior when run in the same order).
Lastly, we also compare the projects containing at least one developer-written or generated flaky test to investigate if generated and developer-written flakiness tends to appear in the same projects.

\subsubsection{RQ2 (Flakiness Suppression)}%
\label{sec:rq2_flakiness_suppression}

To assess the effectiveness of EvoSuite's flakiness suppression, we compare the number of flaky tests generated by \EvoSuiteDefault to those generated by \EvoSuiteFlaky, and to developer-written tests.
For each Java project and each of the three test types, we compute the ratio of flaky to non-flaky tests and use a Wilcoxon signed-rank test~\cite{wilcoxon1945individual}---which is a commonly used, non-parametric paired difference test---to check for statistically relevant differences.
We refrain from using a parametric test, as we found our data to be not normally distributed according to a Shapiro-Wilk test~\cite{shapiro1965analysis}.
Since Pynguin does not offer any optional flakiness suppression mechanisms, we limit our comparison to \EvoSuite.

\subsection{Root Cause Analysis}%
\label{sec:methodology_root_cause_analysis}

In our last research question, we investigate the similarity of de\-vel\-oper-written and generated flaky tests regarding their root causes, which is seen as a core property of flakiness~\cite{Luo2014,Parry2021}.
Like other studies~\cite{Luo2014, Gruber2021}, we categorize the flaky tests' root causes by labeling them manually along established categories.
Namely, we use the amalgamation of root causes collected by Parry et al. in 2021~\cite{Parry2021} as our initial set of pre-defined categories (items 1. to 14. in~\cref{sec:flaky_test_background}).
Since we can automatically detect order-dependency (\OD) through test executions in random orders, we only consider non-order-dependent (\NOD) flaky tests for this step.

\subsubsection{Sampling}%
\label{subsec:sampling}
As we found a total of \num{1740} \NOD flaky tests, we have to take a representative sample to keep the labeling feasible.
To avoid creating a bias towards projects with only a few flaky tests (e.g., by randomly selecting projects), or tests from only a few large projects (e.g., by randomly selecting flaky tests), we combine two sampling strategies:
First, we randomly select one \NOD flaky test from each affected project, regardless of the test type (generated or developer-written), resulting in the \textit{breadth sample}.
Second, we randomly choose \num{21} Java and \num{9} Python projects and sample all their flaky tests (\textit{depth sample}).
These projects are evenly distributed regarding the type---or combination---of flaky tests they contain (Java developer-written, \EvoSuiteDefault, \EvoSuiteFlaky, Python developer-written, Pynguin).
Using this technique, we sampled a total of \num{481} flaky tests (\num{340} Java, \num{141} Python): \num{329} from the breadth sample, \num{122} from the depth sample, and \num{30} selected by both strategies.

\subsubsection{Labeling}%
The manual labeling itself is carried out by four of the authors using the project code, the test code, as well as the test failures (stack trace and error message).
To create a common understanding about what constitutes a certain root cause and to assess if existing root cause categories are applicable to generated flaky tests, we precede the actual labeling with an alignment step:
We randomly choose 50 flaky tests from our sample, which are then classified by all four researchers.
According to Fleiss' Kappa~\cite{fleiss1971measuring}, the four labeling authors have reached an inter-rater reliability of \num{0.41}, which is considered \enquote{good} according to Regier et al.~\cite{regier2013dsm}.
For cases in which the authors disagree, discussions are held, which have resulted in the following adjustments to the set of root causes:
\begin{itemize}
    \item Broaden the category \textit{unordered collection} to also include \textit{unspecified behavior} in general.
    \item Broaden the category \textit{resource leak} to also include \textit{resource unavailability}.
    \item Add category \textit{performance}, which describes tests that fail intermittently due to varying durations of (sequential) processes (example: \cref{lst:performance_devwritten}).
    \item Add category \textit{non-idempotent-outcome} (NIO), which covers self-polluting and self-state-setting tests. This was described by Wei et al.~\cite{wei2022preempting} shortly after the literature survey on which we based our root cause categories~\cite{Parry2021}.
\end{itemize}

After finishing the alignment, the remaining flaky tests in the sample are labeled each by one of the four researchers.

\subsubsection{RQ3 (Root Causes)}%
\label{sec:rq3_root_causes}

We use the labeled root causes to make three main comparisons:
First, the root causes we found in developer-written tests against those found by previous studies.
Second, the root causes of flaky tests generated without flakiness suppression (\EvoSuiteFlaky, Pynguin) against those of developer-written tests of the respective language.
Third, the root causes of tests generated with and without flakiness suppression (\EvoSuiteFlaky vs. \EvoSuiteDefault).

\subsection{Threats to Validity}
\label{sec:threats_to_validity}

\subsubsection{External Validity}
\label{sec:external_validity}

To sample projects for our study, we relied on the Maven Central Repository~\cite{MavenCentralRepository} and PyPI~\cite{pypi}, which are the largest official software repositories of Java and Python.
However, we had to make certain assumptions to keep our setup feasible:
For Java, we only considered projects using Maven and we excluded projects using JUnit~3 due to compatibility issues with more recent versions.
While these design decisions might potentially influence our results, we tried to mitigate this threat by assuming the usage of the predominantly used build automation and testing technologies (Maven and JUnit), which also other studies on test flakiness rely on~\cite{Lam2019,Shi2019a,Alshammari2021}.
For Python, we used an existing dataset of Python projects~\cite{Gruber2021}, which was also used by other researchers to evaluate flakiness detection and debugging techniques~\cite{wei2022preempting,wang2022ipflakies,ahmad2022identifying,gruber2023debugging}.
Nevertheless, we inherit any potentially existing bias in this dataset.

\subsubsection{Construct Validity}
\label{sec:construct_validity}
To detect test flakiness, we executed each test 100 times in a fixed order and 100 times in shuffled orders.
However, some flaky tests have very low failure rates, which might have caused us to underestimate the number of flaky tests in our dataset.
Another potential threat to the construct validity of our study is the search budget used for test generation.
We gave two minutes per Java class for EvoSuite and ten minutes per Python module for Pynguin.
However, allowing more time might have yielded different results.
Choosing a meaningful search budget is a non-trivial issue, especially when setting up experiments that include thousands of projects with various different sizes~\cite{Campos2014}.
To achieve a balance between feasibility, resembling a practical use case, and giving sufficient resources to the tools, we chose our search budgets according to the most commonly used configurations in tool competitions~\cite{Vogl2021,Schweikl2022} or evaluations by the maintainer~\cite{lukasczyk2023empirical}.
We also measured the coverage of the generated tests and found that they yielded a high branch coverage, which indicates that our search budgets were sufficient.

\subsubsection{Internal Validity}
\label{sec:internal_validity}
As we found almost \num{1800} \NOD flaky tests, we had to take a sample before manually labeling their root cause, which might pose a potential threat to the validity of our findings.
To avoid favoring overly large or small projects, we applied a two-fold sampling strategy (see \cref{subsec:sampling}).
Each flaky test was then manually labeled by one of four authors.
This might pose a potential threat, as the authors have different backgrounds and experiences when it comes to root-causing flaky tests.
To mitigate this issue, we created an alignment sample of \num{50} flaky tests that were labeled by all four researchers, and we held discussions about cases in which we disagreed.
In our alignment sample, we reached a \enquote{good} inter-rater reliability, meaning that we were aligned in most of the verdicts given even before starting the alignment.
Furthermore, the root causes we found for developer-written flaky tests match previous studies~\cite{Luo2014,Eck2019}, which increases our confidence in the validity of our other findings.

\section{Results}%
\label{sec:results}

\begin{table*}
    \centering
    \caption{Number of flaky tests found in developer-written and automatically generated tests}
    \label{tab:flaky_test_results}
    \resizebox{\textwidth}{!}{
    \footnotesize
    \begin{tabular}{ccrrrrrrrr}
        \toprule
        Language                & Test Type &
            \multicolumn{2}{|c|}{NOD} &
            \multicolumn{2}{c}{OD} &
            \multicolumn{2}{|c|}{Flaky (NOD + OD)} &
            \multicolumn{2}{c}{All}
            \\[2pt]
                                & & \multicolumn{1}{|c}{\# Tests} & \multicolumn{1}{c|}{\# Projects} & \multicolumn{1}{c}{\# Tests}   & \multicolumn{1}{c}{\# Projects} & \multicolumn{1}{|c}{\# Tests}   & \multicolumn{1}{c|}{\# Projects} & \multicolumn{1}{c}{\# Tests}    & \multicolumn{1}{c}{\# Projects} \\
        \midrule
        \multirow{3}{*}{Java}   & Developer-Written & \num{698} (\SI{0.43}{\percent}) & \num{105} (\SI{5.52}{\percent}) & \num{830} (\SI{0.51}{\percent})   & \num{104} (\SI{5.46}{\percent}) & \num{1528} (\SI{0.94}{\percent}) & \num{161} (\SI{8.46}{\percent}) & \num{163305}  & \multirow{3}{*}{\num{1902}} \\

                                & \EvoSuiteDefault  & \num{175} (\SI{0.06}{\percent}) & \num{43} (\SI{2.26}{\percent})  & \num{1110} (\SI{0.35}{\percent}) & \num{109} (\SI{5.73}{\percent}) & \num{1285} (\SI{0.41}{\percent}) & \num{133} (\SI{6.99}{\percent}) & \num{310193} \\

                                & \EvoSuiteFlaky    & \num{597} (\SI{0.22}{\percent}) & \num{111} (\SI{5.84}{\percent}) & \num{3235} (\SI{1.23}{\percent}) & \num{163} (\SI{8.57}{\percent}) & \num{3832} (\SI{1.45}{\percent}) & \num{228} (\SI{11.9}{\percent}) & \num{264000} \\
        \midrule
        \multirow{2}{*}{Python} & Developer-Written & \num{182} (\SI{0.06}{\percent}) & \num{88}  (\SI{1.98}{\percent}) & \num{1728} (\SI{0.57}{\percent})  & \num{270} (\SI{6.06}{\percent}) & \num{1910} (\SI{0.63}{\percent}) & \num{341} (\SI{7.65}{\percent}) & \num{303711}  & \multirow{2}{*}{\num{4454}} \\

                                & Pynguin           & \num{88} (\SI{0.06}{\percent})  & \num{49}  (\SI{1.10}{\percent}) & \num{925} (\SI{0.67}{\percent})   & \num{183} (\SI{4.11}{\percent}) & \num{1013} (\SI{0.73}{\percent}) & \num{224} (\SI{5.03}{\percent}) & \num{138627} \\
        \bottomrule
    \end{tabular}}
\end{table*}

\subsection{RQ1: Prevalence}%
\label{sec:results_rq1}

\Cref{tab:flaky_test_results} depicts the number of flaky tests we discovered for each language and configuration.
Overall we executed almost 1.2 million tests and discovered \num{9568} flaky tests, roughly two-thirds of them generated ones.
For developer-written tests, we found roughly \SI{0.5}{\percent} to \SI{1}{\percent} of all tests to be flaky, which is similar to previous studies on flakiness in Java~\cite{Lam2019} and Python~\cite{Gruber2021}.
Like them, we also found the ratio between \OD and \NOD flaky tests to be almost even for Java projects, whereas it strongly tilts towards order-dependency for Python projects.

Looking at the flaky tests generated by \EvoSuiteFlaky (\num{3832}) and Pynguin (\num{1013}), we see that for both languages/tools, flakiness is more prevalent in generated than in developer-written tests, relative to the total number of generated/developer-written tests.
In the case of Java, we even see an increase of \SI{54}{\percent} (\SI{0.94}{\percent} to \SI{1.45}{\percent}). %
When distinguishing between order- and non-order-dependent flakiness, we see a strong tendency towards \OD flaky tests (\SI{91}{\percent} of flaky Pynguin tests are \OD, \SI{84}{\percent} of flaky \EvoSuiteFlaky tests are \OD).

\begin{figure}
    \begin{subfigure}[]{0.235\textwidth}
        \includegraphics[width=\linewidth]{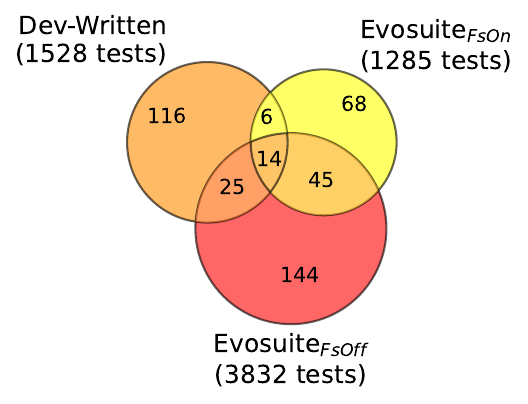}
        \caption{Java}%
        \label{fig:java_venn}
    \end{subfigure}
    \hfill
    \begin{subfigure}[]{0.235\textwidth}
        \includegraphics[width=\linewidth]{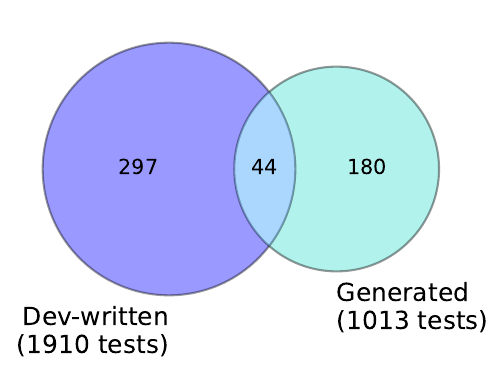}
        \caption{Python}%
        \label{fig:python_venn}
    \end{subfigure}
    \caption{Projects containing flaky tests}
    \label{fig:venn_project_overlap}
\end{figure}

To check if generated flaky tests tend to appear more frequently in projects that already contain developer-written flaky tests, we looked at the sets of projects containing at least one flaky test.
We found \num{224} Python projects containing generated flaky tests and \num{341} projects having developer-written ones.
For Java, \EvoSuiteFlaky produced flaky tests for \num{228} projects, while \num{161} projects contained developer-written flaky tests.
\cref{fig:venn_project_overlap} depicts the overlap between these sets, which is notably small:
Only \SI{17.1}{\percent} of Java and \SI{19.6}{\percent} of Python projects that contain generated flaky tests also contain developer-written flaky tests.

\summary{RQ1: Prevalence}{
    For both Java and Python projects, flakiness is at least as common in generated tests as in developer-written tests. However, it does not appear in the same projects.
    Similar to developer-written tests in Python (but unlike Java tests), the ratio between order-dependent and non-order-dependent flaky tests is leaning strongly towards order-dependency for generated tests.
}

\subsection{RQ2: Flakiness Suppression}
\label{sec:results_suppression}

The second row in~\cref{tab:flaky_test_results} (Java, \EvoSuiteDefault) depicts the amount of flakiness we found among tests generated while using flakiness suppression.
We observe a significant ($p$-value of Wilcoxon test $< 0.001$) reduction in flakiness of \SI{71.7}{\percent} (\SI{1.45}{\percent} to \SI{0.41}{\percent}) compared to \EvoSuiteFlaky and \SI{56.4}{\percent} compared to the developer-written tests.
Among the remaining flaky tests, order-dependency is again far more common than non-order-dependent causes (\SI{86.4}{\percent} of \EvoSuiteDefault flaky tests are \OD).
Looking at projects containing flaky tests (\cref{fig:java_venn}), we see only a minor overlap between \EvoSuiteDefault (\num{133} projects) and the developer-written tests (\num{161} projects), as we found for \EvoSuiteFlaky (\num{228} projects).
When comparing the two EvoSuite configurations, we are surprised to also find only a moderate overlap.
To investigate this observation more deeply, we look at the flaky tests' root causes in \cref{sec:results_rootCauses}.

\summary{RQ2: Flakiness Suppression}{
    EvoSuite's flakiness suppression mechanism is effective: It reduced the number of flaky tests by \SI{71.7}{\percent}, which is considerably lower than the relative number of developer-written flaky tests (\SI{56.4}{\percent} fewer flaky tests).
    The ratio of \NOD and \OD flaky tests remains strongly leaning towards \OD.
}

\subsection{RQ3: Root Causes}%
\label{sec:results_rootCauses}

\setlength{\tabcolsep}{2pt}
\begin{table*}
    \centering
    \caption{Root causes for NOD flaky tests. Cells: number of tests (number of projects)}
    \label{tab:root_causes}
    \footnotesize
    \begin{tabular}{lR{25pt}L{25pt}R{25pt}L{25pt}R{25pt}L{25pt}R{25pt}L{25pt}R{25pt}L{25pt}}
\toprule
{Language}                                                                               & \multicolumn{6}{|c}{\underline{\smash{Java}}}          & \multicolumn{4}{|c}{\underline{\smash{Python}}} \\
{Test Type}                                                                              & \multicolumn{2}{c}{Dev.-Written}               & \multicolumn{2}{c}{\EvoSuiteDefault}               & \multicolumn{2}{c}{\EvoSuiteFlaky}                   & \multicolumn{2}{c}{Dev.-Written}                 & \multicolumn{2}{c}{Pynguin} \\
Total                                                                                    & 170                                            & (106)                                              & 57                                                   & (42)                                             & 113                                            & (102)                                            & 93                                             & (88)                                             & 48                                             & (47)  \\
\midrule
Async Wait                                                                               & \cellcolor[HTML]{08306b}\color[HTML]{f1f1f1}36 & \cellcolor[HTML]{08306b}\color[HTML]{f1f1f1}(23)   & \cellcolor[HTML]{e6f0f9}\color[HTML]{000000}3        & \cellcolor[HTML]{e6f0f9}\color[HTML]{000000}(2)  & \cellcolor[HTML]{cddff1}\color[HTML]{000000}5  & \cellcolor[HTML]{cddff1}\color[HTML]{000000}(4)  & \cellcolor[HTML]{97c6df}\color[HTML]{000000}11 & \cellcolor[HTML]{97c6df}\color[HTML]{000000}(7)  & \cellcolor[HTML]{e5eff9}\color[HTML]{000000}1  & \cellcolor[HTML]{e5eff9}\color[HTML]{000000}(1) \\
Concurrency                                                                              & \cellcolor[HTML]{8dc1dd}\color[HTML]{000000}15 & \cellcolor[HTML]{8dc1dd}\color[HTML]{000000}(14)   & \cellcolor[HTML]{f7fbff}\color[HTML]{000000}0        & \cellcolor[HTML]{f7fbff}\color[HTML]{000000}(0)  & \cellcolor[HTML]{cddff1}\color[HTML]{000000}5  & \cellcolor[HTML]{cddff1}\color[HTML]{000000}(3)  & \cellcolor[HTML]{e2edf8}\color[HTML]{000000}3  & \cellcolor[HTML]{e2edf8}\color[HTML]{000000}(3)  & \cellcolor[HTML]{f7fbff}\color[HTML]{000000}0  & \cellcolor[HTML]{f7fbff}\color[HTML]{000000}(0) \\
Floating Point                                                                           & \cellcolor[HTML]{f7fbff}\color[HTML]{000000}0  & \cellcolor[HTML]{f7fbff}\color[HTML]{000000}(0)    & \cellcolor[HTML]{f7fbff}\color[HTML]{000000}0        & \cellcolor[HTML]{f7fbff}\color[HTML]{000000}(0)  & \cellcolor[HTML]{f7fbff}\color[HTML]{000000}0  & \cellcolor[HTML]{f7fbff}\color[HTML]{000000}(0)  & \cellcolor[HTML]{e9f2fa}\color[HTML]{000000}2  & \cellcolor[HTML]{e9f2fa}\color[HTML]{000000}(2)  & \cellcolor[HTML]{f7fbff}\color[HTML]{000000}0  & \cellcolor[HTML]{f7fbff}\color[HTML]{000000}(0) \\
I/O                                                                                      & \cellcolor[HTML]{c6dbef}\color[HTML]{000000}9  & \cellcolor[HTML]{c6dbef}\color[HTML]{000000}(9)    & \cellcolor[HTML]{f7fbff}\color[HTML]{000000}0        & \cellcolor[HTML]{f7fbff}\color[HTML]{000000}(0)  & \cellcolor[HTML]{f7fbff}\color[HTML]{000000}0  & \cellcolor[HTML]{f7fbff}\color[HTML]{000000}(0)  & \cellcolor[HTML]{cde0f1}\color[HTML]{000000}6  & \cellcolor[HTML]{cde0f1}\color[HTML]{000000}(6)  & \cellcolor[HTML]{bfd8ed}\color[HTML]{000000}3  & \cellcolor[HTML]{bfd8ed}\color[HTML]{000000}(3) \\
Network                                                                                  & \cellcolor[HTML]{08306b}\color[HTML]{f1f1f1}36 & \cellcolor[HTML]{08306b}\color[HTML]{f1f1f1}(9)    & \cellcolor[HTML]{f2f7fd}\color[HTML]{000000}1        & \cellcolor[HTML]{f2f7fd}\color[HTML]{000000}(1)  & \cellcolor[HTML]{56a0ce}\color[HTML]{f1f1f1}13 & \cellcolor[HTML]{56a0ce}\color[HTML]{f1f1f1}(13) & \cellcolor[HTML]{08306b}\color[HTML]{f1f1f1}28 & \cellcolor[HTML]{08306b}\color[HTML]{f1f1f1}(27) & \cellcolor[HTML]{5ca4d0}\color[HTML]{f1f1f1}6  & \cellcolor[HTML]{5ca4d0}\color[HTML]{f1f1f1}(6) \\
NIO                                                                                      & \cellcolor[HTML]{f7fbff}\color[HTML]{000000}0  & \cellcolor[HTML]{f7fbff}\color[HTML]{000000}(0)    & \cellcolor[HTML]{f7fbff}\color[HTML]{000000}0        & \cellcolor[HTML]{f7fbff}\color[HTML]{000000}(0)  & \cellcolor[HTML]{f7fbff}\color[HTML]{000000}0  & \cellcolor[HTML]{f7fbff}\color[HTML]{000000}(0)  & \cellcolor[HTML]{e2edf8}\color[HTML]{000000}3  & \cellcolor[HTML]{e2edf8}\color[HTML]{000000}(3)  & \cellcolor[HTML]{7db8da}\color[HTML]{000000}5  & \cellcolor[HTML]{7db8da}\color[HTML]{000000}(5) \\
OTHER                                                                                    & \cellcolor[HTML]{f7fbff}\color[HTML]{000000}0  & \cellcolor[HTML]{f7fbff}\color[HTML]{000000}(0)    & \cellcolor[HTML]{08306b}\color[HTML]{f1f1f1}34       & \cellcolor[HTML]{08306b}\color[HTML]{f1f1f1}(22) & \cellcolor[HTML]{f7fbff}\color[HTML]{000000}0  & \cellcolor[HTML]{f7fbff}\color[HTML]{000000}(0)  & \cellcolor[HTML]{f7fbff}\color[HTML]{000000}0  & \cellcolor[HTML]{f7fbff}\color[HTML]{000000}(0)  & \cellcolor[HTML]{bfd8ed}\color[HTML]{000000}3  & \cellcolor[HTML]{bfd8ed}\color[HTML]{000000}(3) \\
Performance                                                                              & \cellcolor[HTML]{08468b}\color[HTML]{f1f1f1}33 & \cellcolor[HTML]{08468b}\color[HTML]{f1f1f1}(11)   & \cellcolor[HTML]{f7fbff}\color[HTML]{000000}0        & \cellcolor[HTML]{f7fbff}\color[HTML]{000000}(0)  & \cellcolor[HTML]{b5d4e9}\color[HTML]{000000}7  & \cellcolor[HTML]{b5d4e9}\color[HTML]{000000}(7)  & \cellcolor[HTML]{dbe9f6}\color[HTML]{000000}4  & \cellcolor[HTML]{dbe9f6}\color[HTML]{000000}(4)  & \cellcolor[HTML]{a1cbe2}\color[HTML]{000000}4  & \cellcolor[HTML]{a1cbe2}\color[HTML]{000000}(4) \\
Randomness                                                                               & \cellcolor[HTML]{a3cce3}\color[HTML]{000000}13 & \cellcolor[HTML]{a3cce3}\color[HTML]{000000}(12)   & \cellcolor[HTML]{ebf3fb}\color[HTML]{000000}2        & \cellcolor[HTML]{ebf3fb}\color[HTML]{000000}(2)  & \cellcolor[HTML]{08306b}\color[HTML]{f1f1f1}23 & \cellcolor[HTML]{08306b}\color[HTML]{f1f1f1}(22) & \cellcolor[HTML]{539ecd}\color[HTML]{f1f1f1}16 & \cellcolor[HTML]{539ecd}\color[HTML]{f1f1f1}(16) & \cellcolor[HTML]{08306b}\color[HTML]{f1f1f1}11 & \cellcolor[HTML]{08306b}\color[HTML]{f1f1f1}(11) \\
Resource Leak / Resource Unavailability                                                  & \cellcolor[HTML]{d1e2f3}\color[HTML]{000000}7  & \cellcolor[HTML]{d1e2f3}\color[HTML]{000000}(7)    & \cellcolor[HTML]{ebf3fb}\color[HTML]{000000}2        & \cellcolor[HTML]{ebf3fb}\color[HTML]{000000}(2)  & \cellcolor[HTML]{b5d4e9}\color[HTML]{000000}7  & \cellcolor[HTML]{b5d4e9}\color[HTML]{000000}(6)  & \cellcolor[HTML]{f7fbff}\color[HTML]{000000}0  & \cellcolor[HTML]{f7fbff}\color[HTML]{000000}(0)  & \cellcolor[HTML]{f7fbff}\color[HTML]{000000}0  & \cellcolor[HTML]{f7fbff}\color[HTML]{000000}(0) \\
Test Case Timeout                                                                        & \cellcolor[HTML]{ecf4fb}\color[HTML]{000000}2  & \cellcolor[HTML]{ecf4fb}\color[HTML]{000000}(2)    & \cellcolor[HTML]{dfecf7}\color[HTML]{000000}4        & \cellcolor[HTML]{dfecf7}\color[HTML]{000000}(4)  & \cellcolor[HTML]{08306b}\color[HTML]{f1f1f1}23 & \cellcolor[HTML]{08306b}\color[HTML]{f1f1f1}(18) & \cellcolor[HTML]{f7fbff}\color[HTML]{000000}0  & \cellcolor[HTML]{f7fbff}\color[HTML]{000000}(0)  & \cellcolor[HTML]{f7fbff}\color[HTML]{000000}0  & \cellcolor[HTML]{f7fbff}\color[HTML]{000000}(0) \\
Time                                                                                     & \cellcolor[HTML]{d6e6f4}\color[HTML]{000000}6  & \cellcolor[HTML]{d6e6f4}\color[HTML]{000000}(6)    & \cellcolor[HTML]{ebf3fb}\color[HTML]{000000}2        & \cellcolor[HTML]{ebf3fb}\color[HTML]{000000}(1)  & \cellcolor[HTML]{64a9d3}\color[HTML]{f1f1f1}12 & \cellcolor[HTML]{64a9d3}\color[HTML]{f1f1f1}(12) & \cellcolor[HTML]{dbe9f6}\color[HTML]{000000}4  & \cellcolor[HTML]{dbe9f6}\color[HTML]{000000}(4)  & \cellcolor[HTML]{f7fbff}\color[HTML]{000000}0  & \cellcolor[HTML]{f7fbff}\color[HTML]{000000}(0) \\
Too Restrictive Range                                                                    & \cellcolor[HTML]{ecf4fb}\color[HTML]{000000}2  & \cellcolor[HTML]{ecf4fb}\color[HTML]{000000}(2)    & \cellcolor[HTML]{f7fbff}\color[HTML]{000000}0        & \cellcolor[HTML]{f7fbff}\color[HTML]{000000}(0)  & \cellcolor[HTML]{f7fbff}\color[HTML]{000000}0  & \cellcolor[HTML]{f7fbff}\color[HTML]{000000}(0)  & \cellcolor[HTML]{c6dbef}\color[HTML]{000000}7  & \cellcolor[HTML]{c6dbef}\color[HTML]{000000}(7)  & \cellcolor[HTML]{f7fbff}\color[HTML]{000000}0  & \cellcolor[HTML]{f7fbff}\color[HTML]{000000}(0) \\
UNKNOWN                                                                                  & \cellcolor[HTML]{ccdff1}\color[HTML]{000000}8  & \cellcolor[HTML]{ccdff1}\color[HTML]{000000}(8)    & \cellcolor[HTML]{ebf3fb}\color[HTML]{000000}2        & \cellcolor[HTML]{ebf3fb}\color[HTML]{000000}(2)  & \cellcolor[HTML]{d5e5f4}\color[HTML]{000000}4  & \cellcolor[HTML]{d5e5f4}\color[HTML]{000000}(4)  & \cellcolor[HTML]{c6dbef}\color[HTML]{000000}7  & \cellcolor[HTML]{c6dbef}\color[HTML]{000000}(7)  & \cellcolor[HTML]{5ca4d0}\color[HTML]{f1f1f1}6  & \cellcolor[HTML]{5ca4d0}\color[HTML]{f1f1f1}(5) \\
Unordered Collection / Unspecified Behavior\phantom{x}                                   & \cellcolor[HTML]{e7f0fa}\color[HTML]{000000}3  & \cellcolor[HTML]{e7f0fa}\color[HTML]{000000}(3)    & \cellcolor[HTML]{cfe1f2}\color[HTML]{000000}7        & \cellcolor[HTML]{cfe1f2}\color[HTML]{000000}(6)  & \cellcolor[HTML]{4896c8}\color[HTML]{f1f1f1}14 & \cellcolor[HTML]{4896c8}\color[HTML]{f1f1f1}(13) & \cellcolor[HTML]{e9f2fa}\color[HTML]{000000}2  & \cellcolor[HTML]{e9f2fa}\color[HTML]{000000}(2)  & \cellcolor[HTML]{135fa7}\color[HTML]{f1f1f1}9  & \cellcolor[HTML]{135fa7}\color[HTML]{f1f1f1}(9) \\
\bottomrule
\end{tabular}

\end{table*}
\setlength{\tabcolsep}{6pt}

\Cref{tab:root_causes} depicts the distribution of flakiness root causes that we identified in our sample via manual labeling (\cref{sec:methodology_root_cause_analysis}).

\begin{figure}
\begin{lstlisting}[firstnumber=73,language=Java]
@Test
public void testTimeoutFailExactly() {
    final List mock = Mockachino.mock(ArrayList.class);
    mock.size();
    mock.size();
    runTimeoutTest(Mockachino.verifyExactly(2),200,220,200,500,mock,() -> mock.size());
}`\Suppressnumber`
// ... omitted ...
`\Reactivatenumber{98}`
private void runTimeoutTest(VerifyRangeStart type,int min,int max,int waitTme,int timeout,List mock,Runnable runnable) {

    long t0 = System.currentTimeMillis();
    Executors.newSingleThreadScheduledExecutor().schedule(runnable,waitTime,TimeUnit.MILLISECONDS);
    long t1 = System.currentTimeMillis();
    long margin = t1 - t0;`\Suppressnumber`
    // ... omitted ...  `\Reactivatenumber{106}`
    type.withTimeout(timeout).on(mock).size();`\Suppressnumber`
    // ... omitted ...  `\Reactivatenumber{111}`
    long t2 = System.currentTimeMillis();
    long time = t2 - t1;
    assertTrue(time + " expected at most " + max, time <= max + margin);
}`\Suppressnumber`

<error message="273 expected at most 220" type="junit.framework.AssertionFailedError">`\Reactivatenumber{1}`
\end{lstlisting}
    \caption[Caption]{Developer-written flaky test with root cause \textit{Performance} (project \texttt{krka-mockachino\footnotemark})}%
    \label{lst:performance_devwritten}
\end{figure}
For Java projects, we found \textit{asynchronous waiting} to be a major cause (\SI{21.2}{\percent}) for flakiness in developer-written tests, which corroborates previous studies~\cite{Luo2014,Eck2019}.
However, we also found many flaky tests to be caused by brittle assumptions about the \textit{Performance} (i.e., duration) of sequential processes (\SI{19.4}{\percent}), a root cause that has not previously been described.
\cref{lst:performance_devwritten} shows an example of such a test\footnotetext{\url{https://github.com/krka/mockachino/tree/9bcdda05}}.
The assertion on line~113 is flaky as it assumes that the execution time (line~106) is within a certain range, which is not guaranteed.
For Python projects, the main causes for developer-written flakiness are networking (\SI{30.1}{\percent}) and randomness (\SI{17.2}{\percent}), which was also found by the study from which we sampled our projects~\cite{Gruber2021}.

For flaky tests that were generated without using flakiness suppression (\EvoSuiteFlaky and Pynguin), their root causes fit our pre-defined categories (only few \textit{OTHER} cases), however, the distribution differs:
Generated flaky tests tend to be more commonly caused by \textit{Randomness} ($\sim$\SI{20}{\percent}) and \textit{Unspecified Behavior} (\EvoSuiteFlaky: \SI{12.4}{\percent}, Pynguin: \SI{18.7}{\percent}).
\cref{lst:random_evosuiteflaky} is an example of a test generated by \EvoSuiteFlaky, which is flaky due to randomness.
The test makes an assertion against the value of a random variable that it sampled from a Gaussian distribution using a Box-Muller transform.
\cref{lst:pynguin_random} shows an example of a randomness-related flaky test that was generated by Pynguin for the \texttt{usolitaire} project, which is a terminal solitaire application.
The \texttt{game\_0} object shuffles the deck of cards randomly which causes the \texttt{move\_tableau\_pile(int\_0, bool\_0)} method to arbitrarily pass or fail raising an \texttt{InvalidMove} error.
\cref{lst:python_nod_UC} shows another test that was generated by Pynguin, which is flaky due to the non-deterministic order within a \texttt{frozensets} in Python.

For \EvoSuiteFlaky we also found \textit{Test Case Timeouts} happening frequently (\SI{20.4}{\percent}), which are caused by the 4000ms timeout EvoSuite sets for each test it generates.
Python tests generated by Pynguin, on the other hand, do not exhibit such issues, as it does not set a test case timeout.
One type of flakiness that was generated by Pynguin, but not by EvoSuite, is non-idempotent-outcome (NIO) (\SI{10.4}{\percent}).
\cref{lst:python_nod_NIO} shows a NIO test produced by Pynguin from project \texttt{bl} (BlackEarth core library).
The statement in line 7 tries to create a file, which succeeds for the first run in each iteration (i.e. container), but raises a \texttt{FileExistsError} for every further run.
During the test generation, the \texttt{config\_0.write(str\_0)} operation was most likely executed multiple times, which caused Pynguin to assume that the exception is meant to be thrown, so it created an assertion based on it (line 6).
The test therefore has an inverted failure pattern, where its first execution fails and the following executions pass.
\EvoSuiteDefault doesn't experience such issues, since it uses a virtual file system, however, we also did not find such cases for \EvoSuiteFlaky, where we deactivated this mechanism.

\begin{figure}
    \centering
\begin{lstlisting}[language=Java]
@Test(timeout = 4000)
public void test00()  throws Throwable  {
    RandomJava randomJava0 = new RandomJava();
    randomJava0.gaussian((double) 1057);
    double double0 = randomJava0.gaussian();
    assertEquals(0.8241080392646101, double0, 0.01);
}`\Suppressnumber`

Expected:<0.8241080392646101> but was:<-0.06836772391958745>`\Reactivatenumber{1}`
\end{lstlisting}
    \caption{Randomness-related flaky test generated by \EvoSuiteFlaky (project \texttt{mitchelltech5-jmatharray})}%
    \label{lst:random_evosuiteflaky}
\end{figure}

\begin{figure}
    \centering
\begin{lstlisting}[language=Python]
def test_case_33():
    int_0 = 0
    game_0 = module_0.Game()
    # omitted
    bool_0 = True
    with pytest.raises(module_0.InvalidMove):
        game_0.move_tableau_pile(bool_0, int_0)
    var_0 = game_0.move_tableau_pile(int_0, bool_0)`\Suppressnumber`

E usolitaire.game.InvalidMove`\Reactivatenumber{1}`
\end{lstlisting}
    \caption{Randomness-related flaky test generated by Pynguin \linebreak (project \texttt{usolitaire})}%
    \label{lst:pynguin_random}
\end{figure}

\begin{figure}
    \centering
\begin{lstlisting}[language=Python]
def test_case_54():
  str_0 = ''
  query_0 = module_0.where(str_0)
  # omitted
  query_instance_0 = query_0.all(str_0)
  # omitted
  query_instance_1 = query_0.all(query_instance_0)
  bool_1 = query_instance_1.__call__(dict_0)
  # omitted
  query_instance_2 = query_instance_1.__or__(query_instance_0)
  var_0 = query_instance_2.__repr__()
  assert var_0 == "QueryImpl('or', frozenset({('all', ('',), QueryImpl('all', ('',), '')), ('all', ('',), '')}))"
\end{lstlisting}
    \caption{Unordered collection flakiness generated by Pynguin \linebreak
    (project \texttt{TinyDB})}%
    \label{lst:python_nod_UC}
\end{figure}

\begin{figure}
    \centering
\begin{lstlisting}[language=Python]
def test_case_9():
    none_type_0 = None
    str_0 = '/'
    config_0 = module_0.Config(none_type_0)
    # omitted
    with pytest.raises(FileExistsError):
        config_0.write(str_0)`\Suppressnumber`

E IsADirectoryError: [Errno 21] Is a directory: '/'`\Reactivatenumber{1}`
\end{lstlisting}
    \caption{Non-idempotent-outcome flakiness generated by Pynguin \linebreak (project \texttt{bl})}%
    \label{lst:python_nod_NIO}
\end{figure}

When looking at tests generated while using flakiness suppression (\EvoSuiteDefault), the picture changes drastically:
On one hand, the flakiness suppression vastly reduced the amount of flakiness caused by any known root cause.
On the other hand, we found that the majority (\SI{59.6}{\percent}) of the remaining flaky tests do not fit any known root cause category (\textit{OTHER}).
We inspected these cases in greater detail and found them to be attributable to two causes: \textit{Verifying Expected Exceptions} (18/34) and \textit{StackOverflowErrors} (16/34).

\textit{Verifying Expected Exceptions} describes issues happening when a test case expects a certain exception to be thrown and makes assertions about where (i.e., by which class) the exception was thrown.
In other words, the test case asserts that the top of the stack trace of an expected exception has a certain value.
Such tests can be flaky since a stack trace can change intermittently, even for the same exception.
This is caused by optimizations, namely the just-in-time (JIT) compilation, that might decide at any point during the program execution to compile a frequently executed area in the class to native code, which causes it to no longer appear in the stack trace~\cite{Paleczny2001,Kotzmann2008}.
\cref{lst:thrown_exception_evosuitedefault} shows an example of such a case:
The test is expecting an \texttt{IndexOutOfBoundsException} thrown by \texttt{java.nio.Buffer}.
Sometimes, however, this exception is instead thrown by \texttt{java.nio.HeapByteBuffer}.
This test is flaky due to the default way that the JVM decides to optimize the compilation, where the JIT compilation will compile certain parts of the \texttt{java.nio.Buffer} class to native code, causing it to no longer appear on top of the stack trace (as shown in \cref{fig:tiered_compilation}).
Such optimization-based flakiness does not happen in tests generated by \EvoSuiteFlaky as we updated the \texttt{`No Runtime Dependency'} parameter to \textit{true}, which prevents EvoSuite from generating tests that verify thrown exceptions.
Although keeping it \textit{false} (default) decreases the number of flaky tests of other causes, it also generates new flaky tests that try to verify the class that throws an exception.

Secondly, \EvoSuiteDefault generates flaky tests that produce intermittent \textit{StackOverflowErrors}.
This was also discovered by a previous study~\cite{Fan2019}.
The errors occur consistently when the flakiness suppression is turned off.
\EvoSuiteDefault includes an internal resource threshold---limiting the stack size---to prevent a test case from a \textit{StackOverflowError}, however, the resource checking is non-deterministic and some errors manage to slip through.
Such issues do not occur in \EvoSuiteFlaky because we have disabled the generation of test scaffolding files (first row \cref{tab:evosuiteflaky-parameters}), which include a check to prevent infinite loops in recursive methods.
However, not generating scaffolding files for test classes makes the generated tests more susceptible to traditional causes of flaky tests~\cite{Fraser2018}.

\summary{RQ3: Root Causes}{
    Generated tests are flaky for the same reasons as developer-written ones, however, the distribution among those reasons differs:
    While developer-written flaky tests are often caused by concurrency and networking operations, generated flaky tests tend to be the result of randomness and unspecified behavior.
    When using flakiness suppression, the picture changes vastly, as the majority of remaining flaky tests do not fit any previously described category of flakiness.
    Instead, they are caused by runtime optimizations and EvoSuite-internal resource threshold.
    Notably, both only take effect when certain flakiness suppression mechanisms are activated!
}

\begin{figure}
    \centering

\begin{subfigure}[]{\columnwidth}
    \centering
\begin{lstlisting}[language=Java]
@Test(timeout = 4000)
public void test17()  throws Throwable  {
    Name name0 = new Name();
    int[] intArray0 = new int[1];
    intArray0[0] = (-3152);
    Blob blob0 = new Blob(intArray0);
    SafeBag safeBag0 = null;
    try {
        safeBag0 = new SafeBag(name0, blob0, blob0);
        fail("Expecting exception:IndexOutOfBoundsException");
    } catch(IndexOutOfBoundsException e) {
        verifyException("java.nio.Buffer", e);
    }
}`\Suppressnumber`
Exception was not thrown in java.nio.Buffer but in java.base/java.nio.HeapByteBuffer.get(HeapByteBuffer.java:169): java.lang.IndexOutOfBoundsException: 1`\Reactivatenumber{1}`
\end{lstlisting}
    \caption{Flaky test generated by \EvoSuiteDefault}%
    \label{lst:thrown_exception_evosuitedefault}
\end{subfigure}

    \vskip\baselineskip

\begin{subfigure}[]{\columnwidth}
    \begin{minipage}[t]{0.49\linewidth}
        \begin{framed}
    {\scriptsize
        \begin{Verbatim}[commandchars=\\\{\}]
Exception in thread "main" java.
lang.IndexOutOfBoundsException
\setlength{\fboxsep}{0pt}\colorbox{green!20}{\parbox{\linewidth}{\strut at java.nio.Buffer.checkIndex}}
\setlength{\fboxsep}{0pt}\colorbox{green!20}{\parbox{\linewidth}{\strut  (Buffer.java:743)}}
 at java.nio.HeapByteBuffer.get
  (HeapByteBuffer.java:169)
 ...
 at SafeBag_ESTest.test17
  (SafeBag_ESTest.java:330)
 ...
        \end{Verbatim}
        \textbf{Before JIT compilation}}
        \end{framed}
    \end{minipage}
    \hfill
    \begin{minipage}[t]{0.49\linewidth}
        \begin{framed}
    {\scriptsize
        \begin{Verbatim}[commandchars=\\\{\}]
Exception in thread "main" java.
lang.IndexOutOfBoundsException
\setlength{\fboxsep}{0pt}\colorbox{red!20}{\parbox{\linewidth}{\strut\phantom{at java.nio.Buffer.checkIndex}}}
\setlength{\fboxsep}{0pt}\colorbox{red!20}{\parbox{\linewidth}{\strut\phantom{  (Buffer.java:743)}}}
 at java.nio.HeapByteBuffer.get
  (HeapByteBuffer.java:169)
 ...
 at SafeBag_ESTest.test17
  (SafeBag_ESTest.java:330)
 ...
        \end{Verbatim}
        \textbf{After JIT compilation}}
        \end{framed}
    \end{minipage}
    \vfill
    \caption{Stack traces}
    \label{fig:tiered_compilation}
\end{subfigure}

    \caption{Flakiness due to \textit{Verifying Expected Exceptions} (project \texttt{named-data-jndn})}%
    \label{fig:name}
\end{figure}

\section{Recommendations}
\label{sec:discussion}

\subsection{Maintainers of Test Generation Tools}%

We found EvoSuite's flakiness suppression mechanisms to be very effective and can recommend them to other tools, such as Pynguin, whose rate of flaky tests is currently still higher than the rate of flaky tests in developer-written tests.
Nevertheless, EvoSuite still produces flaky tests, which are mainly caused by (1) \textit{Verifying Expected Exceptions} related to the \texttt{`No Runtime Dependency'} option, and (2) \textit{StackOverflowErrors} caused by scaffolding.
Most notably, both mechanisms are meant---and also accomplish---to prevent traditional causes of flakiness.
We therefore recommend revisiting EvoSuite's implementation of the \texttt{`No Runtime Dependency'} option and the scaffolding mechanisms to eradicate the flakiness these tend to introduce.
Furthermore, we recommend studying and addressing order-dependency in generated tests, as we found high numbers of OD flaky tests for both EvoSuite and Pynguin.

Our dataset~\cite{GeneratedFlakinessDataset} should provide the maintainers of both EvoSuite and Pynguin with a large number of real-world examples that can help to reproduce flakiness in generated tests and serve as an evaluation sample for improved versions.

\subsection{Developers Using Test Generation}%

For developers using EvoSuite, we can highly recommend its flakiness suppression options, which we found to be very effective.
The remaining flaky tests are mostly caused by \textit{Verifying Expected Exceptions} and \textit{StackOverflowErrors}.
The former can be mitigated by disabling the tiered compilation (\texttt{-XX:-TieredCompilation}) flag when executing the tests.
This will prevent the JVM from compiling frequently executed parts of the bytecode into native code (\textbf{JIT} compilation), which however will adversely affect the performance and execution time of other tests.
The flaky \textit{StackOverflowErrors} can be mitigated by removing the \texttt{NonFunctionalRequirementRule} in the test scaffolding file, which however causes the test to fail consistently.
For developers using Pynguin, we recommend setting seeds for random number generators.
This should eliminate randomness-related flakiness, which we found to be the most common individual root cause for flaky Pynguin tests.
However, there is known criticism about using seeding to avoid flakiness~\cite{Dutta2020}.
Developers should also consider order-dependencies between generated tests as a possibility.
In our evaluation, about \SI{4}{\percent} to \SI{6}{\percent} of projects were affected by generated order-dependent flaky tests.

\subsection{Researchers Studying Flaky Tests}%

\label{sec:flaky_tests_automatic_generated_tests_research}
Since generating tests automatically can be done quickly and efficiently, there is a potential for using tools such as \EvoSuite and Pynguin to help in the research on flaky tests.
For example, test generation tools could be used to create training data for machine learning models or to systematically expose non-determinisms in individual target projects.
While we found generated flaky tests to have similar root causes compared to developer-written flaky tests as long as flakiness-suppression mechanisms are turned off (\cref{tab:evosuiteflaky-parameters}), we also found several differences:
First, the root cause distribution differs, as flakiness in generated tests is less likely the cause of concurrency or networking issues, and more often the result of randomness and unspecified behavior (\cref{tab:root_causes}).
Second, we found that projects containing developer-written flaky tests are not particularly likely to also produce generated flaky tests and vice versa (\cref{fig:venn_project_overlap}).

\section{Related work}%
\label{sec:related_work}

Shamshiri et al.~\cite{Shamshiri2015} studied the effectiveness of automatically generated test suites in Java projects.
They applied three automatic test generation tools, Randoop, EvoSuite, and AgitarOne, to a dataset of over 300 faults across five open-source projects, assessing how many bugs the automatically generated tests could detect.
Through this process, they also identified the number of flaky tests that were generated by each of the three tools.
Of the tests generated by Randoop, which uses feedback-directed random test generation, an average of 21\% exhibited non-determinism in their outcomes.
EvoSuite produced flaky tests at an average rate of 3\%.
Only 1\% of the tests generated by the commercial, proprietary tool AgitarOne were flaky.
While our study and theirs both demonstrate that automatic test generation tools are capable of producing flaky tests, there major are differences.
The main objective of our study was to investigate the prevalence and root causes of flaky tests generated by automatic test generation tools.
However, the main objective of the study performed by Shamshiri et al. was to assess the bug-finding capability of automatically generated tests.
As such, we analyzed the prevalence of generated flaky tests in much more detail (see Section \ref{sec:results_rq1}) and went on to categorize their root causes (see Section \ref{sec:results_rootCauses}).
Furthermore, our subject set of \num{1902} Java projects and \num{4454} Python projects is significantly larger than the five Java projects used by Shamshiri et al. in their empirical evaluation.

Paydar et al.~\cite{Paydar2019} examined the prevalence of flaky tests, and other types of problematic tests, generated specifically by Randoop.
They took between 11 and 20 versions of five open-source Java projects and used Randoop to generate regression test suites, which were the main objects of analysis.
Overall, they found that 5\% of the automatically generated test classes were flaky, and on average, 54\% of the test cases within each of these were flaky.
As before, since Paydar et al. were not solely investigating automatically generated flaky tests, they did not examine them to the same level of detail as in our study (for example, they did not consider root causes).
Furthermore, while they are all automatic test generation tools, Randoop is significantly different from EvoSuite and Pynguin in that Randoop is based entirely on random search.

Li et al.~\cite{Li2022} applied automatic test generation to the repair of order-dependent flaky tests.
Their work builds upon the iFixFlakies tool introduced by Shi et al. \cite{Shi2019a}, which uses the statements of existing ``cleaner'' tests to remove the state-pollution left behind by ``polluter'' tests that induce order-dependency in the ``victim'' tests that the tool aims to repair.
A weakness of iFixFlakies is that if no such cleaner test exists in the test suite, the tool will be unable to repair the victim.
The technique introduced by Li et al. aims to address this weakness by applying automatic test generation to generate cleaners such that the victim may be repaired.
Beyond the intersection of flaky tests and automatic test generation, there are no significant similarities between their study and ours.

There have been several previous studies, in which the authors have manually classified the root causes of flaky tests.
Luo et al.~\cite{Luo2014} categorized the causes of the flaky tests repaired by developers in 201 commits across 51 (mostly Java) projects of the Apache Software Foundation.
Vahabzadeh et al.~\cite{Vahabzadeh2015} categorized the causes of 443 bugs in test code (as opposed to in code-under-test), which they mined from the bug repository of the Apache Software Foundation.
They found 51\% of these to be flaky tests, which they went on to subcategorize based on their root cause.
Eck at al.~\cite{Eck2019} asked Mozilla developers to categorize the causes of 200 flaky tests they had previously repaired.
Lam et al.~\cite{Lam2020d} categorized the type of flakiness repaired in 134 pull requests regarding flaky tests in six subject projects that were internal to the Microsoft Corporation.
In spite of much previous work in this area, ours is the first study to categorize the root causes of automatically generated flaky tests.

Flakiness is also an issue in fuzzing, a discipline closely related to automatic test generation:
Nourry et al.~\cite{nourry2023human} surveyed 106 fuzzing practitioners and found that developers are often struggling to reproduce build failures or bugs detected by fuzzing.
Ding et al.~\cite{ding2021empirical} analyzed \num{23907} bugs discovered by OSS-Fuzz~\cite{serebryany2017oss}, of which \SI{13}{\percent} were flaky.
Like our study, they found that timeouts and resource thresholds set by the tool itself (25s and 2.5 GB RAM for OSS-Fuzz~\cite{ossfuzzTimeout}) are major causes of flakiness.

\section{Conclusions}%
\label{sec:conclusions}

Flaky tests are a common and troublesome phenomenon in software testing.
Through this study, we were able to show that flakiness does not only affect developer-written tests, but also automatically generated tests:
We sampled \num{6356} open-source Java and Python projects and generated tests for them using two state-of-the-art test generation frameworks (EvoSuite and Pynguin).
After executing the resulting test suites repeatedly, we found flakiness to be even more common among generated tests than among developer-written ones.
The root causes of this flakiness are similar, however, its distribution differs:
Developer-written flaky tests tend to be caused by concurrency and networking operations, while generated flaky tests are more frequently the result of randomness and unspecified behavior.
For both developer-written and generated flaky tests, order-dependency is a frequent cause, which could be a possible direction for future work.
While flakiness suppression mechanisms are effective in reducing the flakiness rate among generated tests, they also cause other, previously unseen, forms of flakiness.
We hope that our work inspires researchers working on test flakiness and that our insights help maintainers and users of test-generation tools to avoid flakiness in generated tests.
We make all data available~\cite{GeneratedFlakinessDataset}.

\section*{Acknowledgements}\label{sec:acknowledgements}
We thank Stephan Lukasczyk, the creator and maintainer of Pynguin, for his support and advice.
Phil McMinn and Owain Parry are supported by the EPSRC grant "Test FLARE" (EP/X024539/1) and a Meta Testing and Verification award. Owain Parry received additional support from the EPSRC Doctoral Training Partnership with the University of Sheffield (EP/R513313/1).
Gordon Fraser is supported by the DFG project "STUNT" (FR2955/4-1) and by the BMWK project "ANUKI" (50RM2100B).

\balance

\end{document}